# Recent Operation of the FNAL Magnetron H- Ion Source

P.R. Karns[1, a)], D.S. Bollinger[1)], A. Sosa[1)]

[1)]Fermi National Accelerator Laboratory, Box 500, Batavia, Illinois, 60510

[a)] Corresponding author: karns@fnal.gov

**Abstract.** This paper will detail changes in the operational paradigm of the Fermi National Accelerator Laboratory (FNAL) magnetron H$^-$ ion source due to upgrades in the accelerator system. Prior to November of 2012 the H$^-$ ions for High Energy Physics (HEP) experiments were extracted at ~18 keV vertically downward into a 90 degree bending magnet and accelerated through a Cockcroft-Walton accelerating column to 750 keV. Following the upgrade in the fall of 2012 the H$^-$ ions are now directly extracted from a magnetron at 35 keV and accelerated to 750 keV by a Radio Frequency Quadrupole (RFQ). This change in extraction energy as well as the orientation of the ion source required not only a redesign of the ion source, but an updated understanding of its operation at these new values. Discussed in detail are the changes to the ion source timing, arc discharge current, hydrogen gas pressure, and cesium delivery system that were needed to maintain consistent operation at >99% uptime for HEP, with an increased ion source lifetime of over 9 months.

## INTRODUCTION

In November of 2012, Fermi National Accelerator Laboratory (FNAL) began using a Radio Frequency Quadrupole (RFQ) based pre-injector as a replacement for the Cockcroft-Walton (C-W) systems that had provided the H$^-$ beams to the FNAL High Energy Physics (HEP) program since the late 1960's. The C-W systems had become harder to maintain as spare parts had become sparse and experts on the systems retired. The magnetron style ion source that feeds the pre-injector had to be modified as well to better match the RFQ injection parameters. These modifications required the operational parameters of the ion source to be adjusted to the new design.

The FNAL accelerator complex delivers high intensity proton beams to multiple HEP fixed target experiments such as MicroBooNE, MINERvA, and SeaQuest. These experiments have varied beam energies, intensities, and structures that require a flexible H$^-$ ion source that can deliver a 50 – 60 mA beam intensity to the RFQ with a pulse length between 2 µs and 60 µs with a 15 Hz repetition rate. Twin ion sources are operated side by side, with one providing operational H$^-$ beam at any given time, to meet the needs of the HEP program with an expected uptime > 99%. This beam is accelerated to 35 keV as it exits the source through the extractor cone. The H$^-$ beam then travels through a Low Energy Beam Transport (LEBT) that focuses the beam into the RFQ. The RFQ accelerates the beam to 750 keV and feeds the FNAL Linac and later the Booster synchrotron. During Booster injection the H$^-$ beam is stripped of its electrons to allow multiple layers of beam to be injected to increase Booster proton beam intensity. The number of times that this beam is layered is directly related to the H$^-$ beam pulse length that is generated by the ion source. For instance 10 'turns' of beam would require a source pulse length of 22 µs [1].

# EVOLUTION OF THE MAGNETRON ION SOURCE DESIGN

The magnetron H- ion source design used in the C-W preaccelerators [2] is shown in Fig. 1 (a). In this design the magnetron is mounted such that it extracts downward into the extraction channel that accelerates the beam to 18 keV. The magnet poles would then bend the H- ions 90° into the accelerating column that would accelerate the beam to 750 keV suitable for Linac injection.

The magnetron was designed with a long slit for extraction of the H- ions. The cathode within the magnetron evolved to a shape shown in Fig. 1 (b). This oval-shaped cathode has a groove machined into it to provide geometrical focusing of the H- ions to the extraction slit.

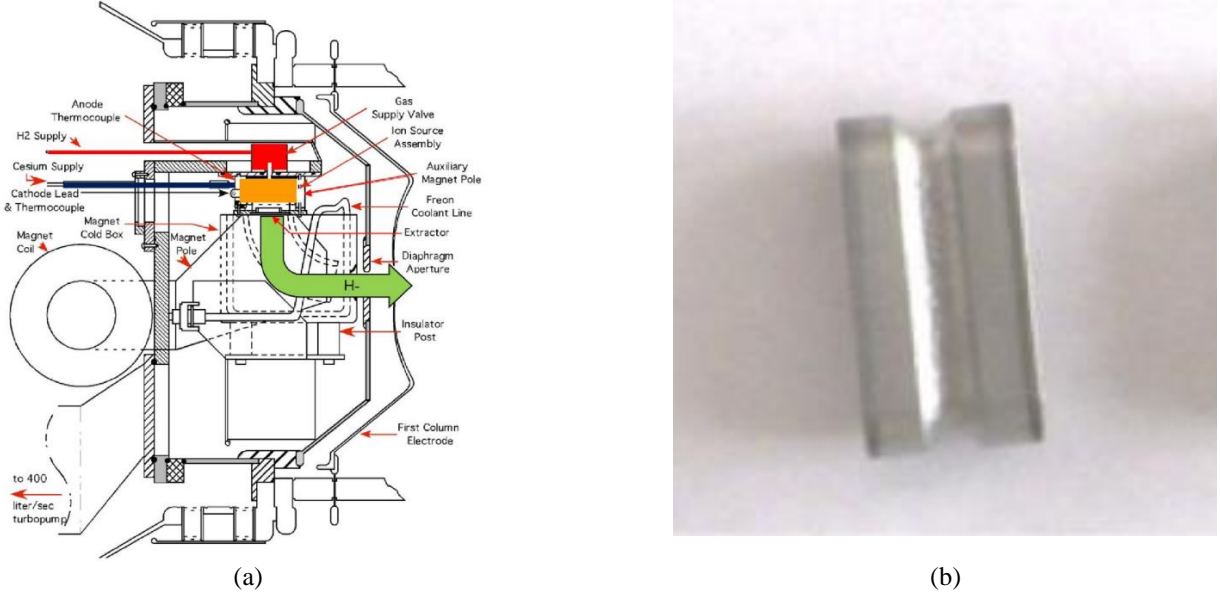

(a) (b)

**FIGURE 1.** The C-W magnetron shown in (a) is mounted to extract downwards through a long slit extractor plate where beam is then bent and focused into the accelerating column. The cathode shape shown in (b) illustrates the groove machined around the entire cathode surface. Original cathode surfaces were flat and evolved into this shape through several iterations.

The ion source was redesigned during the upgrade to the RFQ based preaccelerator [3] as shown in in Fig 2 (a). It was placed such that it now extracted directly into the LEBT that focuses the H- ion beam into the RFQ. It also is the only source of acceleration prior to the RFQ, so the extraction energy was required to match the 35 keV input energy of the RFQ.

In order to provide a round beam to the Linac, the extraction aperture was changed from a long slit to a circular hole. The cathode surface directed towards the circular aperture was changed from a full length groove to a circular dimple as shown in Fig. 2 (b).

It is with these changes in place that new operational parameters had to be found to enable the ion sources to provide the stable, reliable H- ion beams that the FNAL HEP program requires.

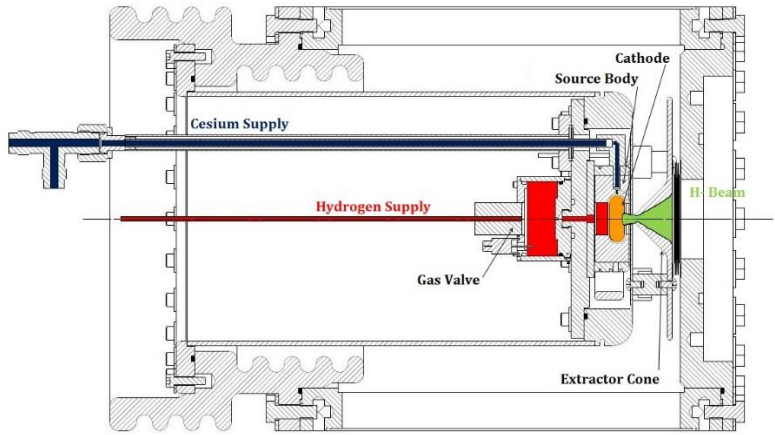 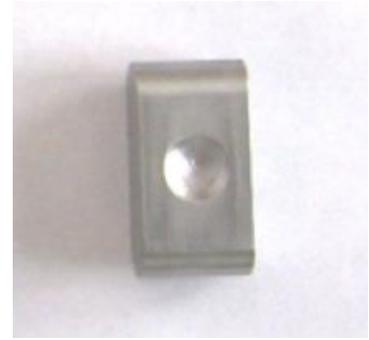

(a)                    (b)

**FIGURE 2.** The RFQ magnetron source (a) is mounted to extract beam directly into the beamline through the cone shaped extractor plate. The cathode shape (b) has a semi-spherical dimple to focus beam into aperture of the extractor cone.

## ION SOURCE TIMING

The former Neutron Therapy Facility (NTF) at FNAL used to require a 62 µs beam pulse for treating patients. This value was used as a maximum beam pulse length that would be requested of the ion source as a typical HEP beam pulse length was <30 µs. The C-W ion sources accommodated this with timing set up such that the extractor pulse started approximately 25 µs before the arc discharge pulse as shown in Fig. 3. The hydrogen gas pulse would start 1 ms before the extractor pulse to allow time for the gas to fill the source. The arc discharge pulse would last for approximately 80 µs and the extractor would pulse for a little over 100 µs. This resulted in a beam flat top of 70-75 µs which would be long enough to meet NTF's requirements.

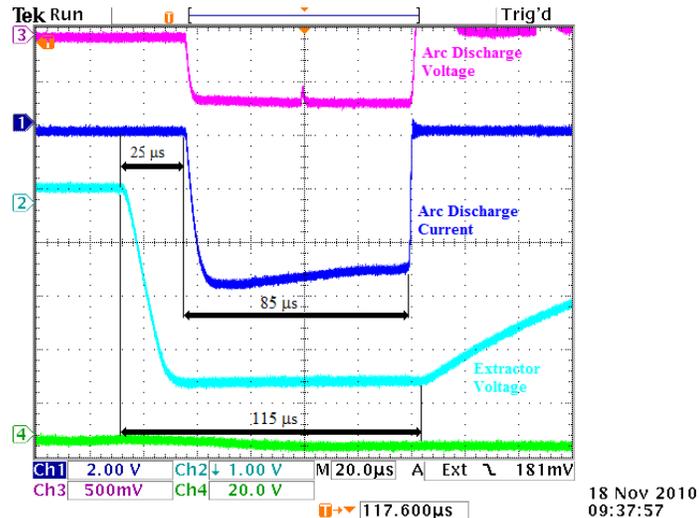

**FIGURE 3.** A scope picture showing the relative source timing of the arc discharge (top and middle) and the extractor (bottom) for the C-W ion sources. The entire arc discharge occurs within the extractor pulse.

These same requirements persisted when the C-W systems were replaced, however some new factors had to be accounted for as well. Part of the design of the RFQ upgrade was to allow for space charge neutralization to take place in the LEBT [3]. Originally it was planned to use Xenon gas to neutralize the H- ion beam which has a neutralization time of 40 µs [4]. This is the time it takes for the passing H- ions to reach a steady state for effective space charge compensation. This beam should not be accelerated by the RFQ.

Initial studies with the beamline utilized the excess Hydrogen that spills from the ion source into the LEBT when it is not converted to H- ions. The characteristic neutralization time for Hydrogen was found to be approximately 60 µs. As a result this 60 µs had to be added to the arc discharge time and the extractor pulse as seen in Fig. 4 to allow for the neutralization to stabilize in the LEBT before beam would be accepted into the RFQ for acceleration. Some time for the beam flattop was saved by shifting the neutralization time earlier utilizing beam extracted from the source above 30 keV. Currently Nitrogen gas is introduced into the LEBT which allows for a shorter neutralization closer to 40 µs similar to Xenon.

The increase of the extractor voltage from 18 kV to 35 kV increased not only the rise time needed for the vacuum tube based extractor pulser, but it also increased the likelihood of the system sparking. This sparking was especially prevalent as the arc discharge pulse began. The sudden addition of free electrons co-extracted from the ion source plasma with the H- ions caused multipactoring between the anode and the extractor cone.

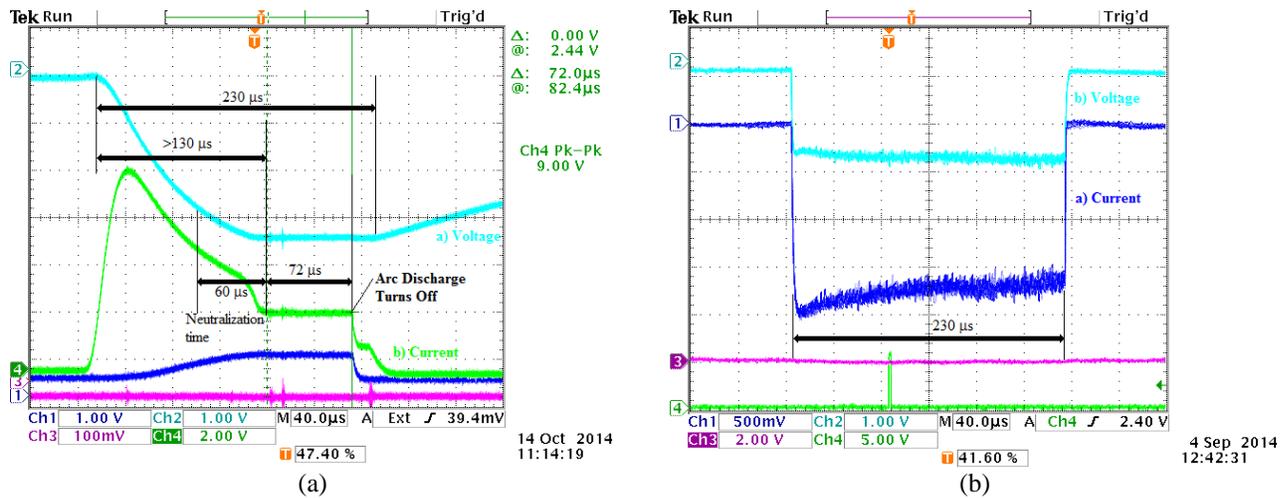

**FIGURE 4.** Scope traces of the new ion source timing. Note that the arc discharge PS and the extractor PS are now isolated due to being at a different potential of 35 kV. The extractor rise time is shown in (a) as well as the beam flattop. The arc discharge in (b) is now 230 µs. Notice that both traces here are shown with 40 µs/div where the scope in Fig. 3 used 20 µs/div.

The arc discharge time was moved earlier than the extractor pulse shown in Fig. 5. This was found to greatly reduce sparking in the anode-extractor gap. This also extended the rise time of the extractor pulser as it now had the impedance of the arc discharge to drive. The rise time of the extractor pulser to 35 kV was now greater than 120 µs seen in Fig. 4 (a). This rise time would slowly lengthen as the vacuum tube in the pulser aged. A tube would last for 3 months before the rise time would become so long that the beam flattop was not long enough to allow for a 62 µs pulse. An arc discharge pulse of 230 µs was now required to maintain a minimum long pulse length of 62 µs as can be seen in Fig. 4 (b). This is nearly a four-fold increase in the arc pulse width compared to the C-W sources.

In November of 2014 the tube based pulser was replaced with a new design that uses solid state IGBT switches [5]. This pulser greatly reduced the extractor rise time to 20 µs, allowing for a beam flattop of 200 µs as shown in Fig 5. The solid state pulser shows no sign of the aging exhibited by the vacuum tube and has been very reliable. This has allowed us recently to reduce the arc discharge time to 100 µs and the extractor pulse width to 110 µs. Early results with this timing have shown no issues.

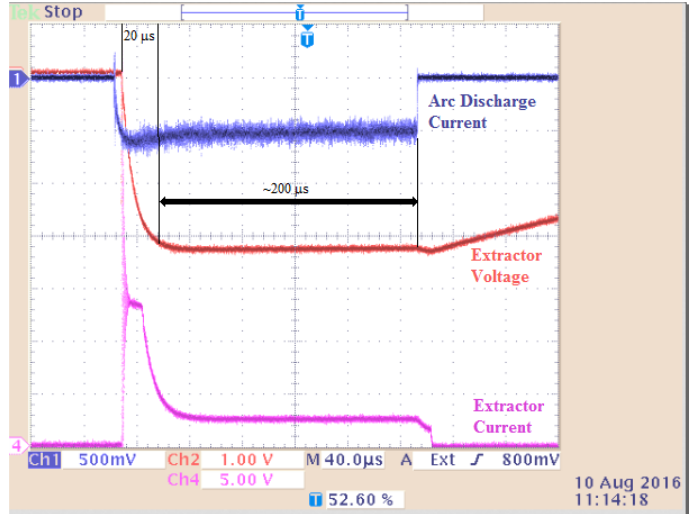

**FIGURE 5.** A scope picture showing the relative source timing of the arc discharge (top) and the solid state extractor (middle and bottom) for the C-W ion sources. The arc discharge begins slightly before the extractor pulse to limit sparking. The solid state switches recovered a large portion of the arc discharge pulse for beam production.

# DECREASED ARC DISCHARGE CURRENT

The increase in the arc discharge pulse width would have been disastrous for the source lifetime had it not been for another consequence of the increased extraction voltage. With the extraction voltage raised to 35 kV, the extractor now pulls a greater quantity of H$^-$ ions from the source plasma [6]. As a result of this, the arc discharge current can be maintained at a much lower value than the C-W sources. A comparison of these values is shown in Table 1. The new design can create 50 – 60 mA of H$^-$ ions at 35 keV with an arc discharge current of 15 A. With the old design the source required an arc discharge current of 55 A to create the same intensity of H$^-$ beam at 18 keV extraction energy.

The lower arc discharge current decreases the damage done to the surface of the cathode done by the incident electrons and Hydrogen species present in the plasma. This allows for the ion source to operate for a longer period before contamination due to cathode erosion degrades the ion source operation. The current design of magnetron ion source has been shown to operate reliably for over 6 months, and over 9 months. As a typical operational period for HEP at FNAL is 9 months, this is a very beneficial feature of the new source design.

**TABLE 1.** Parameter list showing differences between C-W and RFQ era ion sources.

| Parameter | RFQ | C-W |
|---|---|---|
| Arc Discharge Current | 15 A | 45 A |
| Arc Discharge Pulse Width | 230 µs | 80 µs |
| Extractor Voltage | 35 kV | 18 kV |
| Extractor Pulse Width | 230 µs | 115 µs |
| Power Efficiency | 30 mA/kW | 15 mA/kW |
| Duty Factor | 0.38 | 0.13 |
| Vacuum Pressure | 7 x 10$^{-6}$ Torr | 2 x 10$^{-5}$ Torr |
| Source Lifetime | > 9 months | 6 months |

# INCREASED OPERATING SOURCE PRESSURE

Early operation of the ion source was hampered by excessive sparking. This presented a serious problem for operations as the sparking would cause failures of controls equipment, power supplies for ion source components, and eventually the source itself. The original design of the cube within which the ion source rests required two 1200 L/s turbomolecular pumps to reduce the average pressure in the source cube to $2 \times 10^{-6}$ Torr.

Tests were performed with one of the two turbomolecular pumps turned off [7]. The effect of this test can be seen in Fig. 6. When the second pump was turned off, the vacuum pressure jumped from $2 \times 10^{-6}$ Torr to $7 \times 10^{-6}$ Torr. Over the next 18 hours the sparking rate in the source was reduced dramatically. Maintaining a vacuum pressure above $5 \times 10^{-6}$ Torr was found to have a positive effect on the spark rates.

The upper bound of acceptable vacuum pressure in the ion source is set by stripping of the $H^-$ ions. Above a certain pressure the amount of Hydrogen in the plasma will strip the electrons from the ions thus decreasing the extracted ion beam current. Experience with the FNAL magnetron ion source showed this ion stripping to be evident above $9 \times 10^{-6}$ Torr. This gives a typical operating pressure range of $(5-9) \times 10^{-6}$ Torr within the ion source cube.

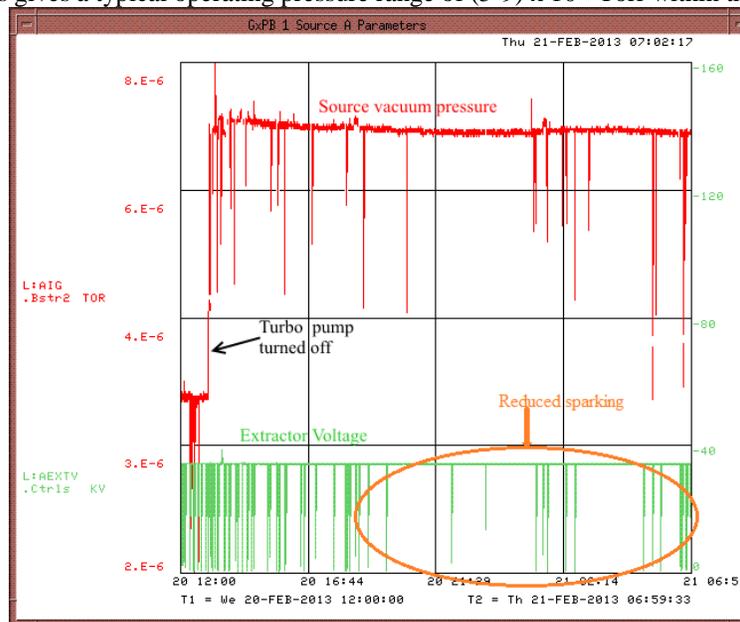

**FIGURE 6.** A plot showing the ion source vacuum pressure (top) and the extractor voltage (bottom). The sparking evident in the repeated drops of the extractor voltage reduces significantly over the next several hours after the pressure was increased.

# DECREASING CESIUM ADMIXTURE

Without an easy way to measure the flow of cesium into the source, the amount of cesium input to the source is determined mostly by the effects that either too little or too much cesium has on the ion source operation. With too little cesium the arc discharge current is impossible to maintain at a high enough level to provide beam for HEP operations. With too much cesium comes the risk of contaminating the ion source and surrounding components as well as increasing the likelihood of sparking in the extraction gap.

Forty years of operating the C-W injectors had defined a suitable level of cesium needed for the ion source to operate efficiently. The cesium delivery system was not appreciably modified in the new source design except to fit the system into the new source can design. It is controlled by a series of heaters attached to each part of the cesium system. The cesium rests in the boiler which is heated to boil the cesium out to the valve. A valve which has its own heater is used to isolate the cesium remaining in the boiler when the ion source is let up to atmosphere for cleaning. A tube connects the valve and boiler to the ion source and has a heating element inside the tube to maintain its temperature.

When first starting to operate the new ion sources, the settings for the cesium heaters were copied to keep the system at similar temperatures. This turned out to provide far too much cesium for the new ion sources that run at a third of the arc discharge current of the old ion sources. Anecdotal evidence from ion source sparks that would trip off the boiler heater led to systematic lowering of the boiler temperature while keeping a close eye on the arc discharge and beam current.

Changes to the cesium boiler can take a long time to have a noticeable effect on the ion source, often taking 16-24 hours to see an appreciable change in the arc discharge current or rate of extractor sparks as shown in in Fig. 7. As a result the boiler temperature was lowered by 3 ˚C and then monitored for the next day. If the source maintained the operational parameters then the temperature was lowered again. This continued until a new low limit for the cesium boiler temperature was found, typically between 110 ˚C and 120 ˚C as compared to > 140 ˚C for the old design.

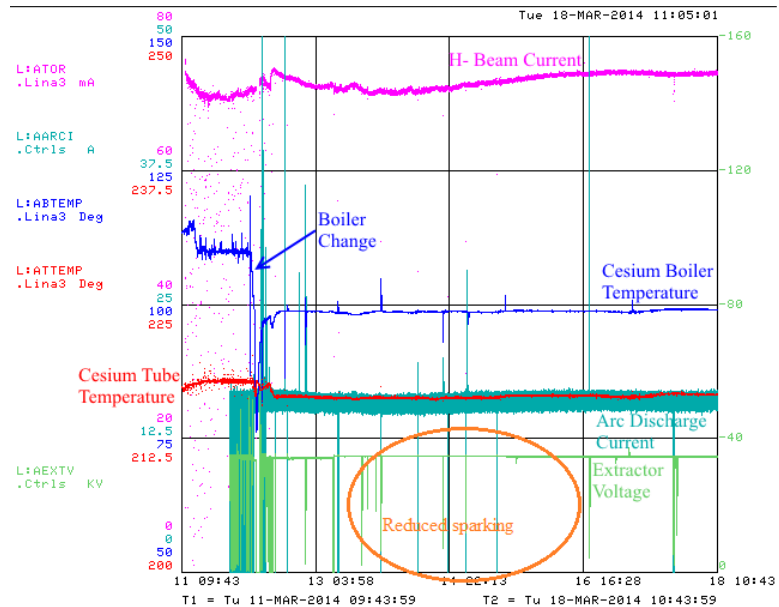

**FIGURE 7.** A plot showing the effect of reducing cesium in the source by lowering the boiler temperature. Sparking in the source as seen in the drops in extractor voltage reduces within 24-48 hours of lowering the boiler temperature by ~12 ˚C.

This also led to a change in the way that cesium is used when starting an ion source after cleaning [7]. The old method was to increase the cesium to higher than operational values and then lower the temperature after the arc discharge was stable above 15 A. This would typically lead to sparking early during an ion source lifetime. The new method was to start with the cesium system deliberately low and allow a small arc discharge typically around 5 A to become stable. The cesium boiler temperate was then raised slowly while keeping the other source parameters constant until the arc discharge reached the target value.

These changes in the cesium usage have also extended the lifetime of a boiler before it would have to be cleaned and refilled. A new cesium boiler begins operations with a 5 g vial of cesium. With the old sources typically this would last 400-450 days. With the lower usage ion sources a vial has lasted over 640 days.

## CONCLUSION

The change from the C-W to the RFQ preaccelerator was mandated by the changing times and limited resources for the older C-W systems. The requirements of the new RFQ system to meet the needs of the FNAL HEP program have been met while fine tuning the ion sources to run at higher ion source vacuum pressure, lower arc discharge current, and lower cesium consumption. These changes have allowed for more stable H$^-$ beam above 50 mA with a longer ion source lifetime over 9 months flexible enough to meet the varied demands of the program.

## ACKNOWLEDGEMENTS

This research was supported by Fermi Research Alliance, LLC under Contract No. De-AC02-07CH11359 with the United States Department of Energy. We would like to thank Andrew Feld and Ken Koch for their technical expertise in working on the ion sources as well as Chuck Schmidt for all his help in understanding the details of the systems.